\documentstyle[11pt,newpasp,twoside]{article}
\def\edcomment#1{\iffalse\marginpar{\raggedright\sl#1\/}\else\relax\fi}
\reversemarginpar

\begin{document}
\title{WINGS: a Wide--field Imaging Nearby Galaxy clusters Survey}
\author{G.Fasano, D.Bettoni, C.Marmo, E.Pignatelli, B.M.Poggianti}
\affil{Padova Astronomical Observatory, Vicolo dell'Osservatorio 5, I-35122}
\author{M.Moles}
\affil{CSIC, C/ Serrano 123, E-28006 Madrid}
\author{P.Kj{\ae}rgaard}
\affil{Copenhagen University Observatory, Juliane Maries Vej 30, DK-2100}

\begin{abstract}
We present a two-band, wide-field imaging survey of an X--ray selected
sample of 78 clusters in the redshift range z=0.03--0.07. The aim of
the project is to provide the astronomical community with a complete
set of homogeneous, CCD--based, surface photometry and morphological
data of the nearby cluster galaxies located within 1.5Mpc from the
cluster center.
\end{abstract}

\section{Motivation}

In all statistical studies concerning photometric and morphological
evolution of galaxies in clusters, the only reference sample available
in the nearby universe is the historical one of Dressler (1980) who
lists the positions, the visual morphological classifications and the
approximate magnitudes of galaxies in 55 clusters at z=0.011--0.066.
It is obvious that a local reference sample with quality adequate to
the modern technologies is missing and is crucial both for studying
the morphological content of nearby clusters in a systematic way and
for setting the zero-point for evolutionary studies.

This project is aimed at filling in this gap through a complete CCD
surface photometry (in the B and V bands) of galaxies in a
well-defined sample of low--z clusters. 

\section{Cluster sample and status of observations}

The growing availability of efficient and reliable Wide--Field Imaging
CCD Cameras (WFICs) has opened for the first time the possibility of
gathering in a reasonable time a large amount of CCD data on galaxies
in nearby clusters. We have exploited this opportunity in both the
north (INT, La Palma) and the south (ESO--2.2) hemispheres, by taking
observations of clusters selected from an essentially complete X-Ray
flux limited sample of Abell clusters (XBACs, Ebeling et al. 1996)
compiled from ROSAT All--Sky Survey data. The redshift range
(0.04$<$z$<$0.07) is the only selection criterion applied, resulting
in a total sample of 78 clusters (36 in the northern hemisphere and 42
in the southern one) over a broad range in X-Ray luminosity. Since now
a total of 14 observing nights have been already devoted to this
project, resulting in the complete imaging of 61 clusters.

\section{Immediate objectives and techniques}

The data will be used to produce, for each cluster of the sample,
the following outputs:

-- a deep (V$_{lim}\sim$22.5; B$_{lim}\sim$23.5) photometric catalog
of galaxies containing coordinates, integrated V and B magnitudes,
concentration index, rough ellipticity and position angle estimates of
each object;

-- a surface photometry catalog, relative to a sub-sample of
bright/large enough galaxies of the previous, deep list. In this
catalog we will include the whole photometric and morphological
information extracted from the luminosity and geometrical profiles
(color profile, isophotal twisting, disky/boxy c$_4$ profile,
Bulge/Disk decomposition, etc..), as well as the global parameters
(effective radius and average surface brightness, total magnitude,
Sersic's index, morphological type, etc..) of each galaxy;

The reduction procedures are clearly crucial to achieve these
scientific objectives. In particular, automatic pipelines for data
reduction and surface photometry are required, together with a fast
computing capability.

The deep catalogs will be produced running SExtractor (Bertin and
Arnout, 1996) onto the co-added frames in the two filters, whereas the
surface photometry catalogs will be produced by using a tool for
Galaxy Automatic Surface PHOTometry in wide and/or deep fields
(GASPHOT, Pignatelli and Fasano 1999) we are presently developing at
the Padova Observatory. It consists of four main tools:

a) STARPROF produces a careful (space varying) representation of the
PSF profile;

b) SEXISOPH exploits some capabilities of SExtractor to quickly
produce luminosity and geometrical profiles of the galaxies;

c) GALPROF fits the equivalent luminosity profiles of galaxies by
using both a Sersic law and a three component ($r^{1/4}$ + exp + PSF)
profile convolved with the proper PSF and produces unbiased estimates
of total magnitudes, effective radii, Sersic's indices, etc..;

d) MORPHOT exploits some characteristic features of the luminosity and
geometrical profiles to estimate the morphological type of individual
galaxies.

\end{document}